# Substantial drag reduction in turbulent flow using liquid-infused surfaces

Tyler Van Buren† AND Alexander J. Smits†

Experiments are presented that demonstrate how liquid-infused surfaces can reduce turbulent drag significantly in Taylor-Couette flow. The test liquid was water, and the test surface was composed of square microscopic grooves measuring 100 $\mu$m to 800 $\mu$m, filled with alkane liquids with viscosities from 0.3 to 1.4 times that of water. We achieve drag reduction exceeding 35%, four times higher than previously reported for liquid-infused surfaces in turbulent flow. The level of drag reduction increased with viscosity ratio, groove width, fluid area fraction, and Reynolds number. The optimum groove width was given by $w^+ \approx 35$.

## 1. Introduction

Much of the recent research on drag reduction in turbulent wall-bounded flows has focused on modifying the zero-slip wall boundary condition. In particular, Superhydrophobic Surfaces (SHS) have been shown to reduce drag in laminar and turbulent water flows by up to 50% (Srinivasan *et al.* 2013; Ou *et al.* 2004; Martell *et al.* 2009; Park *et al.* 2013; Daniello *et al.* 2009; Park *et al.* 2014). In SHS, the surface is treated so that air can be trapped in asperities formed on the surface. This can be achieved by using a patterned surface finish (such as small grooves, micro-pillars, or other regular structures), or a heterogenous surface finish obtained by (for example) spraying or chemical treatments, that are functionalized to have large contact angles for water, so that they preferentially hold air when submerged.

The case of SHS using simple transverse grooves is illustrated in figure 1 (the grooves in this study are longitudinal, but the concepts noted here are similar). For the flow in the regions where the water is directly in contact with the solid surface, the no-slip condition holds, and as a consequence the shear at the surface is large. In the regions where the water is flowing over a cavity filled with air, which has a viscosity about 50 times smaller than that of water, the interface is able to slip, and so the shear is reduced. When the local shear stress is integrated over the entire surface, a marked decrease in drag is possible.

However, SHS suffer from some significant drawbacks. The integrity of the water/air interface depends on surface tension, and so turbulent pressure fluctuations, or an increase in hydrostatic pressure associated with a change in depth, can lead to failure of the interface (Bocquet & Lauga 2011; Poetes *et al.* 2010). Similarly, a high level of shear can cause a breakdown of the interface (Samaha *et al.* 2012). Once the cavities have been flooded with water, the surface asperities present themselves as roughness elements, causing an overall drag increase rather than the intended drag decrease (Aljallis *et al.* 2013).

† Mechanical and Aerospace Engineering, Princeton University, Princeton NJ 08544, USA

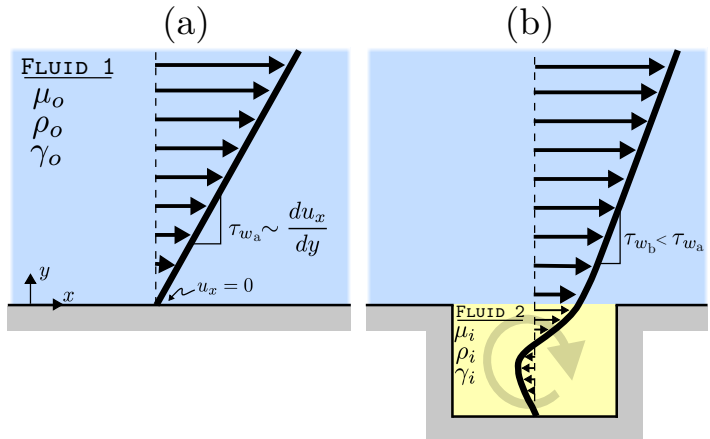

Figure 1: Schematic of the near-wall velocity behavior of single fluid system (a) and a two fluid system (b). The presence of the second fluid creates a recirculation region reducing the near-wall shear and drag induced by the flow.

In a new approach, we replace the air trapped in the surface features with a liquid that is immiscible in water, to form a Liquid-Infused Surface (LIS) (Wong *et al.* 2011). Such surfaces are inherently robust to pressure changes, and so they avoid one of the major weaknesses of SHS. However, given that the infused liquid will most likely have a considerably higher viscosity compared to air, it is not obvious that such surfaces will help to reduce drag. In that respect, drag reduction has been shown to be possible in laminar flows (Solomon *et al.* 2014), and in turbulent flows (Rosenberg *et al.* 2016), although the turbulent drag reduction was only about 10%, considerably less than what seems to be possible with SHS.

We now present new results of drag reduction using LIS that are as high as 35%, a level that is comparable to the drag reduction that can be achieved using SHS in the same experiment. Indications are that even higher levels may be possible in the future.

There are four primary parameters governing the performance of LIS that stem from interrelationships between viscous stress, inertial force, buoyancy force, and interfacial tension. The first is the viscosity ratio, $N = \mu_o/\mu_i$, where $\mu_o$ and $\mu_i$ are the viscosities of the external and infused fluid, respectively. Typical values for LIS are of $\mathcal{O}(1)$, whereas for SHS it is approximately 50. Higher viscosity ratios result in a higher potential drag reduction (Schönecker *et al.* 2014). The second is the Reynolds number, which is the ratio of the inertial forces to the viscous forces. The Reynolds number is given by $Re = \rho_o U d/\mu_o$, where $\rho_o$ is the density of the external fluid, $U$ is the characteristic flow velocity, and $d$ is a characteristic length scale. In our case, the characteristic velocity is the outer surface velocity and the length scale is the gap size in the Taylor-Couette apparatus. High Reynolds numbers indicate a turbulent flow (in our case the maximum Reynolds number is 10,500, which is turbulent for Taylor-Couette flows).

The other two parameters are important in maintaining the infused liquid/water interface. One is the Bond number, which is the ratio of the buoyancy force to the capillary force. When the infused and external fluids have different densities, the resulting buoyancy force can deplete the surface features by overcoming the capillary force holding the fluid in. The Bond number is given by $Bo = (\rho_o - \rho_i)gw^2/\gamma$, where $w$ is the characteristic length scale of the surface asperities (in our study is the groove width), $\gamma$ is the interfacial tension, and $\rho_0$ and $\rho_i$ are the densities of the external and infused fluid, respectively.

| Case | Infused fluid | $N$ | $\rho_i$ [kg/m$^3$] | $Bo \times 10^{-2}$ | $We \times 10^{-1}$ |
|------|---------------|-----|---------------------|---------------------|---------------------|
| SHS  | Air           | 47.8 | 1.23  | 1.4—10.8 | 0.3—2.43 |
| LIS  | Hexane        | 3.0  | 655   | 0.68—5.4 | 0.44—3.5 |
| LIS  | Heptane       | 2.3  | 684   | 0.62—5.0 | 0.44—3.5 |
| LIS  | Octane        | 1.8  | 703   | 0.58—4.7 | 0.44—3.5 |
| LIS  | Decane        | 1.1  | 730   | 0.53—4.2 | 0.44—3.5 |
| LIS  | Undecane      | 0.8  | 740   | 0.51—5.1 | 0.44—3.5 |
| LIS  | Dodecane      | 0.7  | 750   | 0.49—3.9 | 0.44—3.5 |

Table 1: Infused fluid properties with water as the bulk fluid. Ranges of Weber ($We$) and Bond ($Bo$) numbers are given corresponding to groove sizes of $w = 100$ $\mu$m and 800 $\mu$m and the highest Reynolds number, $Re = 10,500$. The viscosity ratio $N = \mu_o/\mu_i$.

The other parameter is the ratio of the inertial force to the capillary force, that is, the Weber number, given by $We = \rho_o U^2 w/\gamma$. When the Weber number is high, the inertia force can overcome the capillary force and allow the infused liquid to escape. For a robust LIS application, the infused and external liquids need to be immiscible, the infused liquid needs to preferentially wet the surface, and the Weber and Bond numbers need to be small to avoid failure.

Along with the properties of the impregnated fluid, the surface geometry significantly impacts drag reduction. Surfaces with larger fluid area fraction, $a = A_i/A_t$ (where $A_i$ is the surface area of the exposed infused fluid and $A_t$ is the total surface area), have a higher drag reduction potential. Also, surfaces with equal fluid area fraction but larger features tend to reduce drag more (Fu *et al.* 2016) because fewer areas that transition from a solid to a liquid boundary will be present, and so regions where locally high stresses can occur are reduced. The primary reason we achieve more substantial drag reduction than in Rosenberg *et al.* (2016) is by employing larger surface features. It should be noted that there is a lesser understood limit to the size of the features – larger wetted areas might be more susceptible to failure inducing interfacial instabilities (Mohammadi & Smits 2016).

Here, we use surface features made by longitudinal square grooves with widths ranging from $w = 100$ $\mu$m to 800 $\mu$m. The majority of experiments were performed at a fluid area ratio $a = 50\%$, though one case is shown for $a = 85\%$ (with $w = 400\mu$m). The external fluid is water, and the infused-liquid is a set of alkanes, with viscosity ratios ranging from $N = 0.7$ to 3. We also consider the classical superhydrophobic (air-infused) case, $N = 50$, for purposes of comparison.

## 2. Experiment

The experiment was conducted in a Taylor-Couette rheometer facility (Brabender Rheotron), illustrated in figure 2, equipped with a stainless steel outer cylinder (radius $R_o = 28$ mm) that rotates around the treated stationary inner cylinder ($R_i = 26$ mm). The gap between the two cylinders is $d = R_o - R_i = 2$ mm $\pm 0.013$ mm, and the cylinder height is $H = 80$ mm so that $H/d = 40$. An acrylic ring, located 500 $\mu$m above the top of the inner cylinder, minimizes the rise of the free surface of the water during rotation. As the outer cylinder rotates at angular velocity $\omega$, it induces a characteristic surface velocity of $U = \omega R_o$ and a torque ($T$) on the stationary cylinder, monitored via a 25 gram load cell (LCM systems UFI) through a 25 mm lever arm. The accuracy of the load cell is $< \pm 0.3\%$ full-scale, which for our experiments corresponds to a torque accuracy of $\pm 1\text{-}2\%$, with better accuracy at higher Reynolds numbers. The load cell output signal was conditioned by a low-pass Butterworth filter and amplifier (Krohn-Hite Corporation), and each experiment is run three times independently to ensure repeatability and gain a sense of uncertainty which yields an average standard deviation of 2% in torque measurements.

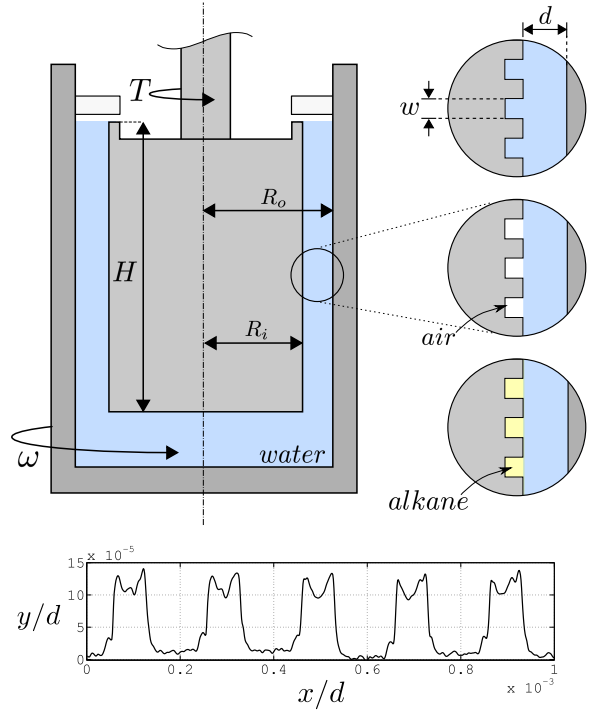

Figure 2: Experimental setup (above) of the Taylor-Couette facility. Longitudinal grooves are tested with water (baseline), air (superhydrophobic), or alkane oils (liquid-infused). Sample confocal microscope measurements (below) of the smallest groove size, $w = 100$ $\mu$m are shown.

The cylinder is first dip-coated in the infused liquid, then immediately submerged in the test liquid (water) in order to limit the exposure of the infused liquid to air — the lower viscosity alkanes are especially volatile. Because the alkane oils are less dense than water, any excess oil floats to the free-surface, which is then removed to minimize impact on the experiment. It is possible that a small nanolayer of oil remains on the oleophilic outer surface of the cylinder after the dip coating and buoyancy driven drainage. This volume of fluid will be small with respect to the groove volume, but simulations have indicated that this nanolayer helps keep surface asperities submerged as the fluid-fluid interface deforms, thereby improving drag reduction (Arenas *et al.* 2016).

For liquid infusion, we created test pieces with surface features that were initially preferentially wetting to the surrounding water. A lathe was used to machine the square surface grooves, which ranged from $w = 100$ $\mu$m to 800 $\mu$m wide, and the machine accuracy was confirmed using a confocal microscope (Olympus LEXT OLS4000). To chemically treat the surface to attract alkanes, the following steps were followed: (i) plasma cleaning to remove all surface organic compounds; (ii) ultrasonic cleaning to remove remaining non-organic impurities, such as machining oil or metal chips; (iii) Boehmetization (boiled in water), following the procedure of (Kim *et al.* 2013), which creates a nano-textured oxide layer that enhances the surface hydrophobicity; and (iv) immersion in a solution of 1% trichlorosilane (OTS) and 99% heptane for 4-8 hours, which covers the surface in a mono-layer of polymers that result in a surface which is superhydrophobic, but attracting to alkane oils.

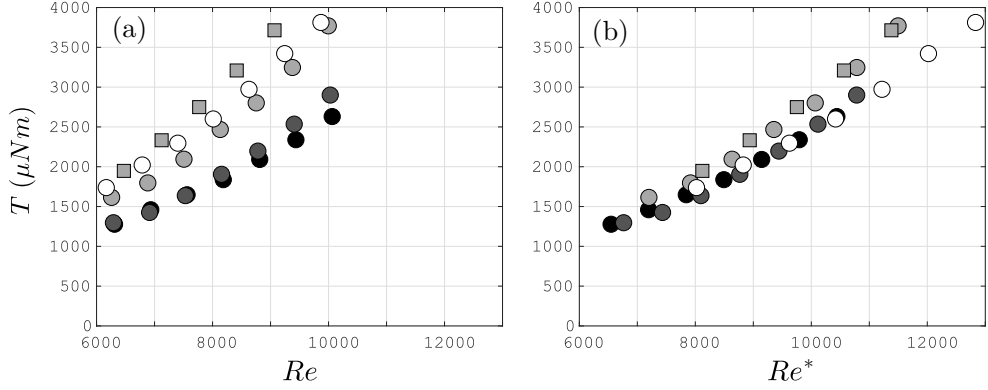

Figure 3: Water-filled groove cylinder torque with Reynolds number based on (a) gap height $d$, and (b) effective gap height $d_{eff}$. Groove sizes $w = 100$ $\mu$m (black), 200 $\mu$m (dark gray), 400 $\mu$m (light gray), and 800 $\mu$m (white); fluid area fractions $a = 0.5$ (circular symbols) and 0.85 (square symbols).

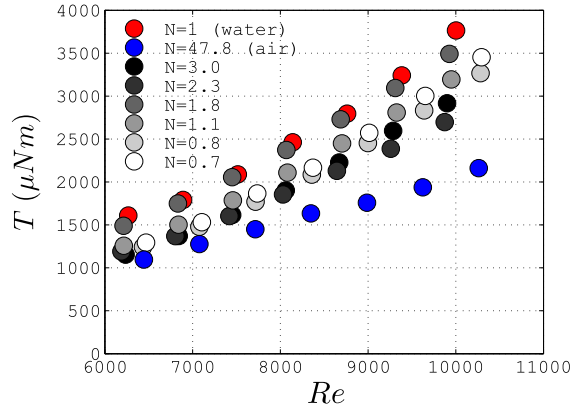

Figure 4: Cylinder torque as a function of Reynolds number for the baseline (no surface infusion), air (superhydrophobic), and liquid infused cases represented by their viscosity ratio, $N$. Here, $a = 0.5$.

## 3. Results

First, we consider the baseline case, where the surface is hydrophilic and water fills the grooves. Figure 3(a) shows that as the groove size increases, the torque at a given Reynolds number increases. Here, $Re = \rho_o U d/\mu_o$, where $d = R_o - R_i$. However, in this study some of the larger groove sizes are not negligible to the gap height. We can define an effective inner cylinder radius based on an equivalent volume of test fluid, where $R_{eff}^2 = R_i^2 - a\left(R_i^2 - (R_i - w)^2\right)$, and hence an effective gap width $d_{eff} = d + \mathcal{M}(R_i - R_{eff})$, where $\mathcal{M} = 1.5$ is an empirically determined constant that tunes the importance of the grooves (for $\mathcal{M} = 1$ the effective gap width is equal to its average value based on the equivalent volume of fluid). Figure 3(b) shows the torque $T$ for the water filled grooves as they vary with Reynolds numbers based on the effective gap height, $Re^* = \rho_o U d_{eff}/\mu_o$. The effective Reynolds number collapses the data reasonably well, indicating that the change in torque for the different groove sizes is mostly due to the change in the effective Reynolds number. For any particular groove size, therefore, we can use the water-filled groove case as the baseline for determining drag reduction.

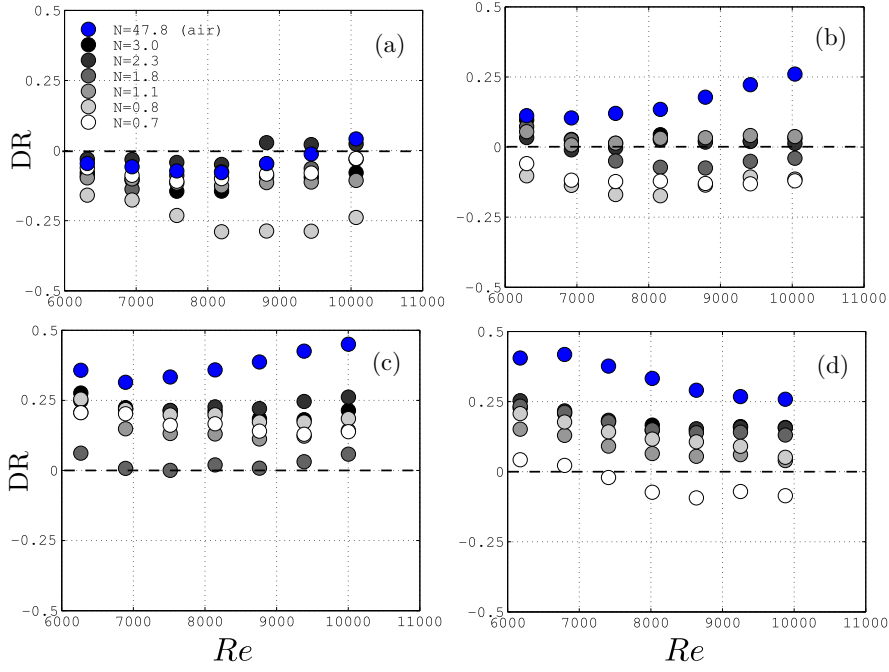

Figure 5: Drag reduction with respect to the baseline (no surface treatment) for air (superhydrophobic) and liquid infused cases represented by their viscosity ratio, $N$. Groove sizes (a) $w = 100$ $\mu$m, (b) 200 $\mu$m, (c) 400 $\mu$m, and (d) 800 $\mu$m. Here, $a = 0.5$.

Figure 4 shows a sample set of torque measurements for a groove width of $w = 400$ $\mu$m. Three test cases were studied over a range of Reynolds numbers: the baseline (water-filled, $N = 1$) case, the superhydrophobic (air-filled, $N = 47.8$) case, and the liquid-infused (alkane-filled, $3 \geq N \geq 0.7$) case. Generally, the torque increases with Reynolds number because at higher spin rates there is more friction drag on the inner cylinder. For this geometry, replacing the water within the grooves with air or alkanes decreases the cylinder torque at a fixed Reynolds number, with the superhydrophobic case producing the lowest drag.

The drag reduction $DR$ is defined as the change in torque $T$ with respect to the baseline torque $T_0$, so that $DR = (T_0 - T)/T_0$. The results are shown in figure 5 for groove widths of $w = 100$ $\mu$m to 800 $\mu$m. For all groove sizes there is a general trend that higher viscosity ratios results in a larger drag reduction, as expected (Schönecker *et al.* 2014; Fu *et al.*

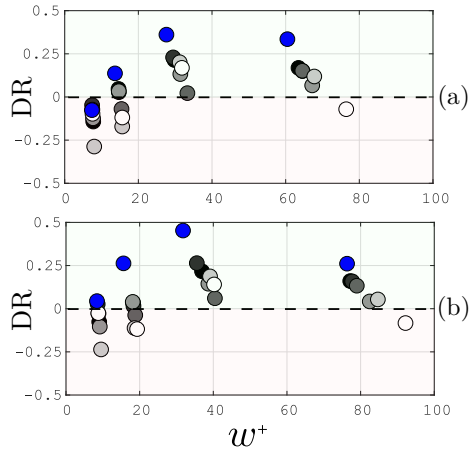

Figure 6: Groove width impact on drag reduction for fixed Reynolds numbers of (a) $Re_d = 8,000$ and (b) $10,000$ for air (superhydrophobic) and liquid infused cases. Here, $a = 0.5$. Symbols as in figure 5.

2016). This behavior is not entirely consistent, because there are some higher viscosity oils that produce larger drag reduction than the lower viscosity cases, particularly the $w = 400$ $\mu$m case. This is counter-intuitive, and the most likely cause of this behavior is surface failure like the appearance of instabilities on the interface (Mohammadi & Smits 2016), although this suggestion remains to be confirmed. Unfortunately, the nature of these Taylor-Couette measurements does not allow us to assess surface failure during the experiment.

Figure 6 shows the drag reduction at each groove size for Reynolds numbers $Re = 8,000$ and $10,000$. As suggested by Park et al. (2014)(Park *et al.* 2014), we scale the groove width with the viscous length scale of the flow such that $w^+ = w\nu^{-1}\sqrt{\tau_w/\rho}$ where $\tau_w = T/2\pi H R_i^2$ is the wall stress and $\nu$ is the kinematic viscosity of water. The largest drag reduction is at $w^+ \approx 35$, approaching 45% for the air-infused case and greater than 25% for the liquid-infused case. The location of the peak drag reduction is in good agreement with superhydrophobic results (Park *et al.* 2014). Larger groove sizes produce larger drag reduction, as indicated by (Park *et al.* 2013, 2014). However, for the largest groove size at the higher Reynolds numbers the drag reduction decreases with increasing Reynolds number. Visual inspection of these cases showed that water had wetted a portion of the surface, indicating a partial surface failure under these conditions. The results from these Reynolds numbers are consistent with our entire measurement range, though it should be noted that this range is limited.

As suggested by Fu *et al.* (2016), increasing the fluid area fraction $a$ increase the level of drag reduction. Figure 5 shows results for $a = 0.5$, and figure 7 shows results for $a = 0.85$. At lower Reynolds numbers, we see that increasing $a$ increases the drag reduction from about 25% to up to 35% for the liquid infused cases at this groove width (400 $\mu$m).

Similar drag reduction levels have been reproduced in the MIT Taylor-Couette apparatus (Rajappan & Mckinley 2017). In the MIT apparatus (most recently described in Saranadhi *et al.* 2016) the inner cylinder rotates while the outer one is stationary (the opposite is true for the Princeton apparatus), the gap size $d = 12.7$ mm and $H/d = 3.3$. When comparing to a *smooth* cylinder as the baseline case, they see a drag reduction of up to 15% for heptane infused grooves, $w = 200$ $\mu$m ($w^+ = 50$) and $a = 0.5$, at Reynolds numbers up to $Re = 40,000$. This result compares well with the 18% drag reduction shown in this study for the most similar geometry and flow condition ($w = 400$ $\mu$m, $w^+ = 40$, and $a = 0.5$).

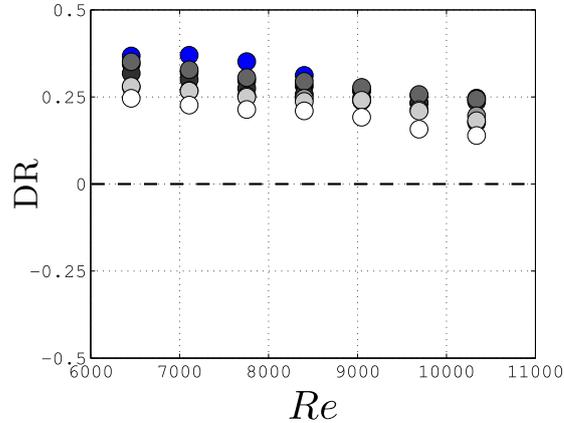

Figure 7: Drag reduction for fluid area fraction $a$=0.85 with respect to the baseline (no surface treatment) for air (superhydrophobic) and liquid infused cases represented by their viscosity ratio, $N$ ($w = 400$ $\mu$m). Symbols as in figure 5.

## 4. Conclusions

We demonstrated drag reduction in turbulent Taylor-Couette flow up to 35% with liquid-infused surfaces and 45% for superhydrophobic surfaces. Generally, the level of drag reduction increased with viscosity ratio, groove width, fluid area fraction, and Reynolds number. Some cases did not follow expected trends, as when higher viscosity fluids produced larger drag reductions, which is most likely due to surface failure. The optimum groove width was given by $w^+ \approx 35$. For the largest grooves, the surfaces mostly failed during testing, likely because the buoyancy and inertial forces were large enough to overcome the capillary force keeping the fluid inside the grooves.

Many supporting studies are still needed, specifically in boundary layers, with finite-length grooves, and at higher Reynolds numbers to verify that these surfaces will perform under more realistic conditions.

We thank Prof. Wong of Pennsylvania State University and Prof. Aizenberg of Harvard University for their advice on chemical surface treatments. This work is funded by the Office of Naval Research through MURI Grant Nos. N00014-12-1-0875 and N00014-12-1-0962 (Program Manager Dr. Ki-Han Kim).

REFERENCES

Aljallis, E., Sarshar, M. A., Datla, R., Sikka, V., Jones, A. & Choi, C.H. 2013 Experimental study of skin friction drag reduction on superhydrophobic flat plates in high reynolds number boundary layer flow. *Physics of Fluids* **25** (2), 025103.

Arenas, I., Bernardini, M., Iungo, G. V. & Leonardi, S. 2016 Turbulent drag reduction over super-hydrophobic and liquid infused surfaces: dependence on the dynamics of the interface. In *31st Symposium on Naval Hydrodynamics*. ONR.

Bocquet, L. & Lauga, E. 2011 A smooth future? *Nature Materials* **10** (5), 334–337.

Daniello, R.J., Waterhouse, N.E. & Rothstein, J.P. 2009 Drag reduction in turbulent flows over superhydrophobic surfaces. *Physics of Fluids* **21** (8), 085103.

Fu, M. K., Mohammadi, A., Van Buren, T., Stone, H. A., Smits, A. J., Hultmark, M., Arenas, A. & Leonardi, S. 2016 Understanding the effects of finite viscosity in super-hydrophobic and liquid infused surface drag reduction. In *31st Symposium on Naval Hydrodynamics*. ONR.

Kim, P., Kreder, M.J., Alvarenga, J. & Aizenberg, J. 2013 Hierarchical or not? effect of the length scale and hierarchy of the surface roughness on omniphobicity of lubricant-infused substrates. *Nano Letters* **13** (4), 1793–1799.

Martell, M.B., Perot, J.B. & Rothstein, J.P. 2009 Direct numerical simulations of turbulent flows over superhydrophobic surfaces. *Journal of Fluid Mechanics* **620** (1), 31–41.

Mohammadi, A. & Smits, A. J. 2016 Stability of two-immiscible-fluid systems: a review of canonical plane parallel flows. *Journal of Fluids Engineering* **138** (10), 100803.

Ou, J., Perot, B. & Rothstein, J.P. 2004 Laminar drag reduction in microchannels using ultrahydrophobic surfaces. *Physics of Fluids* **16** (12), 4635–4643.

Park, H., Park, H. & Kim, J. 2013 A numerical study of the effects of superhydrophobic surface on skin-friction drag in turbulent channel flow. *Physics of Fluids* **25** (11), 110815.

Park, H., Sun, G. & Kim, J. 2014 Superhydrophobic turbulent drag reduction as a function of surface grating parameters. *Journal of Fluid Mechanics* **747**, 722–734.

Poetes, R, Holtzmann, K., Franze, K. & Steiner, U. 2010 Metastable underwater superhydrophobicity. *Physical Review Letters* **105** (16), 166104.

Rajappan, A. & Mckinley, G. H. 2017 personal communication.

Rosenberg, B.J., Van Buren, T., Fu, M.K. & Smits, A.J. 2016 Turbulent drag reduction over air-and liquid-impregnated surfaces. *Physics of Fluids* **28** (1), 015103.

Samaha, M.A., Tafreshi, H.V. & Gad-el Hak, M. 2012 Influence of flow on longevity of superhydrophobic coatings. *Langmuir* **28** (25), 9759–9766.

Saranadhi, D., Chen, D., Kleingartner, J. A., Srinivasan, S., Cohen, R. E. & McKinley, G. H. 2016 Sustained drag reduction in a turbulent flow using a low-temperature leidenfrost surface. *Science Advances* **2** (10), e1600686.

Schönecker, C., Baier, T. & Hardt, S. 2014 Influence of the enclosed fluid on the flow over a microstructured surface in the cassie state. *Journal of Fluid Mechanics* **740**, 168–195.

Solomon, B.R., Khalil, K.S. & Varanasi, K.K. 2014 Drag reduction using lubricant-impregnated surfaces in viscous laminar flow. *Langmuir* **30** (36), 10970–10976.

Srinivasan, S., Choi, W., Park, K.C.L, Chhatre, S.S., Cohen, R.E. & McKinley, G.H. 2013 Drag reduction for viscous laminar flow on spray-coated non-wetting surfaces. *Soft Matter* **9** (24), 5691–5702.

Wong, T.S., Kang, S. H., Tang, S. K.Y., Smythe, E. J., Hatton, B. D., Grinthal, A. & Aizenberg, J. 2011 Bioinspired self-repairing slippery surfaces with pressure-stable omniphobicity. *Nature* **477** (7365), 443–447.